\documentclass[a4paper,twocolumn,keywords,synopsis]{iucr}
\usepackage[latin9]{inputenc}
\usepackage{amsmath}
\usepackage{amssymb}
\usepackage{graphicx}
\usepackage{esint}
\makeatletter

     \paperprodcode{a000000}      % Replace with production code if known
     \paperref{xx9999}            % Replace xx9999 with reference code if known
     \papertype{FA}               % Indicate type of article
     \paperlang{english}          % Can be english, french, german or russian
     \journalcode{A}             % Indicate the journal to which submitted
     \journalyr{2020}
     \journalreceived{\relax}
     \journalaccepted{\relax}
     \journalonline{\relax}

\makeatother

\begin{document}
\title{Small-angle X-ray scattering from GaN nanowires on Si(111): facet
truncation rods, facet roughness, and Porod's law}
\shorttitle{SAXS from GaN nanowires}
\author[a]{Vladimir M.}{Kaganer}
\author[b]{Oleg V.}{Konovalov}
\author[a,c]{Sergio}{Fernández-Garrido}
\aff[a]{Paul-Drude-Institut für Festkörperelektronik, Leibniz-Institut im
Forschungsverbund Berlin e.\,V., Hausvogteiplatz 5--7, 10117 Berlin,
Germany}
\aff[b]{European Synchrotron Radiation Facility, 71 avenue des Martyrs, 38043
Grenoble, France}
\aff[c]{Grupo de electrónica y semiconductores, Dpto.\ Física Aplicada,
Universidad Autónoma de Madrid, C/ Francisco Tomás y Valiente 7, 28049
Madrid, Spain}
\shortauthor{Kaganer, Konovalov, and Fernández-Garrido}
\maketitle
\begin{synopsis}
Small-angle X-ray scattering intensity from GaN nanowires on Si(111)
depends on the orientation of the side facets with respect to the
incident beam. This reminiscence of the truncation rod scattering
gives rise to a deviation from Porod's law. A roughness of just 3--4
atomic steps per a micron long side facet notably changes the intensity
curves. 
\end{synopsis}
\begin{abstract}
Small-angle X-ray scattering from GaN nanowires grown on Si(111) is
studied experimentally and modeled by means of Monte Carlo simulations.
It is shown that the scattering intensity at large wave vectors does
not follow Porod's law $I(q)\propto q^{-4}$. The intensity depends
on the orientation of the side facets with respect to the incident
X-ray beam. It is maximum when the scattering vector is directed along
a facet normal, as a reminiscence of the surface truncation rod scattering.
At large wave vectors $q$, the scattering intensity is found to be
decreased by surface roughness. A root mean square roughness of 0.9~nm,
which is the height of just 3--4 atomic steps per micron long facet,
already gives rise to a strong intensity reduction. 
\end{abstract}

\section{Introduction}

GaN nanowires (NWs) spontaneously form in plasma-assisted molecular
beam epitaxy (PA-MBE) on various substrates at elevated temperatures
under excess of N \cite{garrido09,garrido12}. In contrast to the
vapor--liquid--solid (VLS) growth approach followed to synthesize
the majority of semiconductor NWs, PA-MBE growth of GaN NWs takes
place without a metal particle on the top \cite{ristic08}. Advantages
of the spontaneous formation are the absence of contamination from
foreign metal particles and the possibility to fabricate axial heterostructures
with sharp interfaces by alternating the supply of different elements.

GaN NWs on Si(111), which is the most common substrate, grow in dense
ensembles ($\gtrsim10{}^{10}$~cm$^{-2}$) and initially possess
radii of tens of nm as well as broad radius and length distributions
\cite{consonni2013pss}. As they grow in length, they bundle together,
which results in an enlargement of the radii at their top parts \cite{kaganer16bundling}.
Regarding their epitaxial orientation, GaN NWs on Si(111) possess
a 3--5\,$^{\circ}$ wide distribution of their orientations with
respect to both the substrate normal (tilt) and the in-plane crystallographic
orientation of the substrate (twist) \cite{jenichen11}.

For dense NW ensembles on Si(111), the radius distribution can be
obtained from the analysis of top-view scanning electron micrographs
\cite{brandt14}. However, this method provides the radius distribution
of only the top part of the NWs, which notably differs from the radius
distribution at their bottom part because of NW bundling. In addition,
the use of scanning electron micrographs for the statistical analysis
of the NW radii becomes much more laborious for NW ensembles with
low densities as those formed on TiN \cite{treeck18}, since when
the magnification of the scanning electron micrographs is chosen to
quantify the NW diameters, only a few NWs fall into the field of view.

Small-angle X-ray scattering is potentially better suited than scanning
electron microscopy for the determination of the radius distribution
of GaN NWs ensembles grown on Si(111) because it probes the entire
NW volume. From the standpoint of small-angle scattering, GaN NWs
are long hexagonal prisms with a substantial distribution of their
cross-sectional sizes and orientations. Since these NWs are, on average,
aligned along the substrate surface normal, the incident X-ray beam
is to be directed at a grazing incidence to the substrate surface.
Grazing incidence small angle X-ray scattering (GISAXS) has been employed
to study Si \cite{david08,buttard13}, GaAs \cite{mariager07}, and
InAs \cite{eymery07,eymery09,mariager09} NWs grown by the VLS growth
mechanism with Au nanoparticles at their tops. Unlike spontaneously
formed GaN NWs, NW ensembles prepared by VLS are characterized by
very narrow distributions of the NW sizes and orientations. The scattering
intensity from such NW ensembles possesses the same features as the
scattering intensity from a single NW: it exhibits oscillations due
to the interference caused by reflections at opposite facets and a
pronounced intensity dependence on the facet orientation.

In the case of GaN NWs, despite the potential advantages of GISAXS
to assess the distribution of NW radii, we are not aware of any GISAXS
study. The closest report is the work of \citeasnoun{horak08},
who performed an in-plane X-ray diffraction study of GaN NWs using
a laboratory diffractometer. Their analysis implies the absence of
strain in the NWs. If so, the NW diameters can be obtained from $\omega/2\theta$
scans in the same way as it can be done in GISAXS. However, this analysis
cannot be applied to dense arrays of GaN NWs, which are inhomogeneously
strained as a result of NW bundling \cite{jenichen11,kaganer12NWstrain,garrido14,kaganer16xray}.
We do not discuss here other X-ray diffraction studies of NWs devoted
to the determination of strain and composition since they are out
of the scope of the present work.

The aim of the present paper is to develop the approaches required
for the analysis of GaN NW arrays by GISAXS using dense NW ensembles
grown on Si(111) as a model example. Since GaN NWs are faceted crystals
(their side facets are $\left\{ 1\bar{1}00\right\} $ planes), we
expected that the GISAXS intensity at large wave vectors follows Porod's
law. Porod's law \cite{porod51,debye57} states that, at large wave
vectors $q$, the small-angle scattering intensity $I(q)$ from particles
with sharp boundaries (i.\,e., possessing an abrupt change of the
electron density at the surface) follows a universal asymptotic law
$I(q)\propto q^{-4}$. \citeasnoun{sinha88} pointed out a common
origin of Porod's law in small-angle scattering and Fresnel's law
for reflection from flat surfaces. Namely, the scattering intensity
from a planar surface in the $xy$ plane is proportional to $q_{z}^{-2}\delta(q_{x})\delta(q_{y})$.
An average over random orientations of the plane gives rise to the
$q^{-4}$ law just because the delta function $\delta(q)$ has a dimensionality
of $q^{-1}$. \citeasnoun{sinha88} performed an explicit calculation
of the orientational average. Deviations from Porod's law are caused
by fractality or the roughness of the surfaces in porous media \cite{bale84,wong88,sinha89}.

In this paper, we show that the GISAXS intensity from GaN NWs at large
wave vectors depends on the azimuthal orientation of the NW ensemble
with respect to the incident X-ray beam. The intensity is maximum
when the scattering vector is directed along the facet normal, and
minimum when the scattering vector is parallel to the facet. In other
words, the azimuthal dependence of the GISAXS intensity reveals the
facet truncation rods. They are well established in X-ray diffraction
from nanoparticles \cite{renaud09} and stem from crystal truncation
rods from planar surfaces \cite{Robinson1986a,robinson92}. We also
show that the large-$q$ intensity reveals the roughness of the side
facets of the GaN NWs. We determine a root mean squared (rms) roughness
of about 0.9~nm, corresponding to the height of a few atomic steps
on a micron long NW sidewall facet.

\section{Experiment}

\label{sec:Experiment}

For the present study, we have selected three samples with different
NW lengths from the series A studied by \citeasnoun{kaganer16bundling}.
The GaN NWs were synthesized in a molecular beam epitaxy system equipped
with a solid-source effusion cell for Ga and a radio-frequency N$_{2}$
plasma source for generating active N. The samples were grown on Si$(111)$
substrates, which were preliminarily etched in diluted HF (5$\%$),
outgassed above 900\,$^{\circ}$C for 30~min to remove any residual
Si$_{x}$O$_{y}$ from the surface, and exposed to the N plasma for
$10$~min. The substrate growth temperature was approximately 800\,$^{\circ}$C,
as measured with an optical pyrometer. The Ga and N fluxes, calibrated
by determining the thickness of GaN films grown under N- and Ga-rich
conditions \cite{heying00}, were $0.29$ and $0.75$ monolayers per
second, respectively. The growth time is the only parameter that was
varied among the samples to obtain ensembles of NWs with different
lengths.

%-------------------------------------------------------------------------------
\begin{figure}
\includegraphics[width=1\columnwidth]{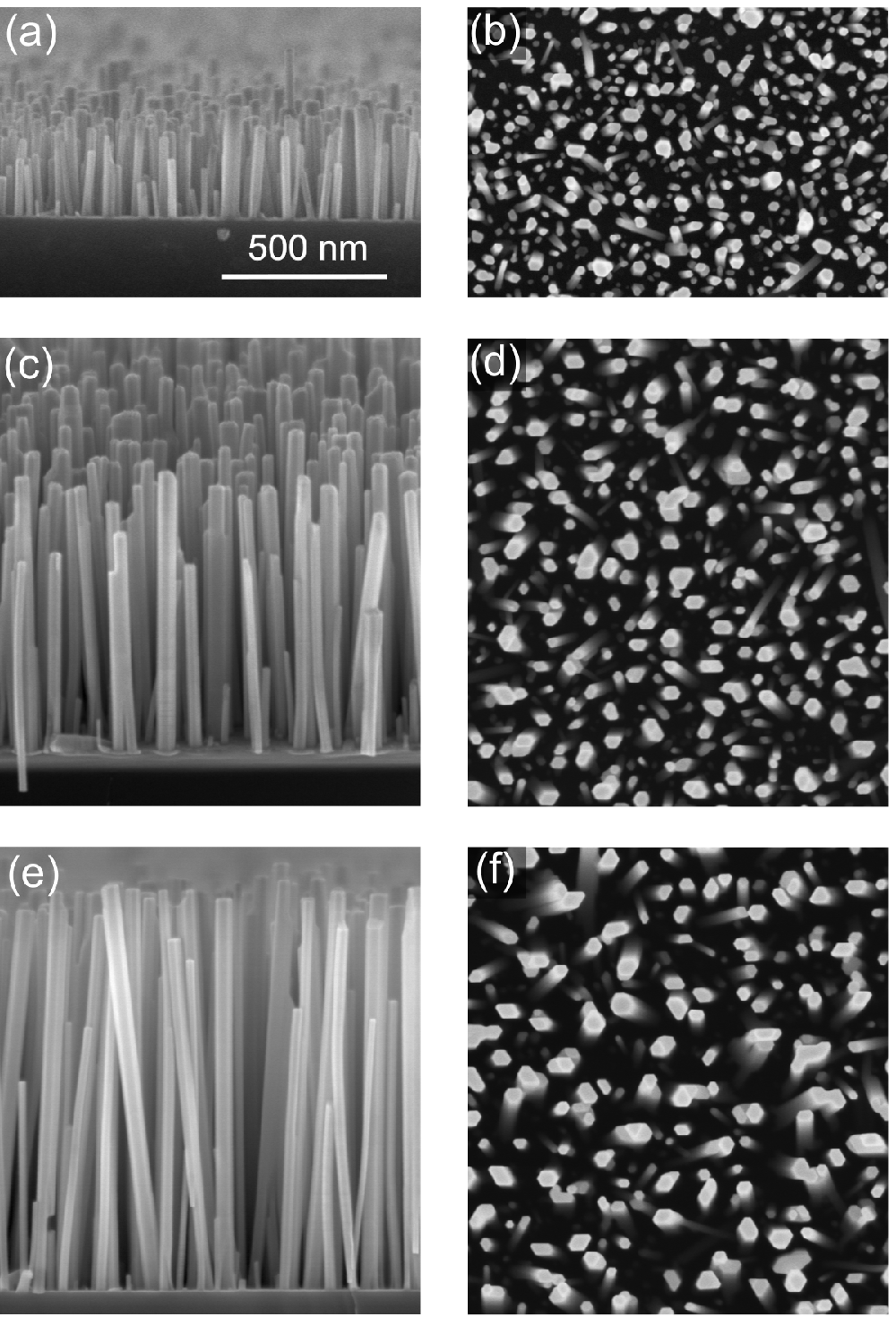}

\caption{Bird's eye view (left column) and top-view (right column) scanning
electron micrographs of samples 1 (a,b), 2 (c,d), and 3 (e,f). The
average NW lengths are 230, 650, and 985~nm, respectively. The scale
bar in (a) is applicable to all micrographs.}

\label{fig:SEMimages} 
\end{figure}

%-------------------------------------------------------------------------------

Figure \ref{fig:SEMimages} presents scanning electron micrographs
of samples 1--3. Sample 1 corresponds to the end of the NW nucleation
process. The NW density is $3.5\times10^{10}$~cm$^{-2}$, while
the average length and diameter of the NWs are 230 and 22~nm, respectively.
The NWs are mostly uncoalesced hexagonal prisms. Samples 2 and 3 display
the further growth of the NWs, with average NW lengths of 650~nm
for sample 2 and 985~nm for sample 3. The average NW diameters, as
determined by the analysis of the top-view micrographs (right column
of Fig.~\ref{fig:SEMimages}), increase with increasing NW lengths,
while the NW density decreases. \citeasnoun{kaganer16bundling}
showed that the increase in the diameter is a result of NW bundling,
rather than their radial growth. A decisive proof of the absence of
radial growth comes from the measurement of the fraction of the total
area that is covered by NWs. The area fraction covered by the NWs,
derived from the top-view micrographs shown in the right column of
Fig.~\ref{fig:SEMimages}, does not change from one sample to the
other and remains always at 20\%.

The distribution of the NW orientations was determined with a laboratory
X-ray diffractometer. We have measured the full width at half-maximum
(FWHM) of the GaN 0002 reflection to determine the tilt range
with respect to the substrate surface normal and the GaN $1\bar{1}00$
reflection to determine the twist range with respect to the
in-plane orientation of the substrate. The FWHM of the tilt distribution
is found to decrease with the NW length from 5.1\,$^{\circ}$ for
sample 1 to 4.0\,$^{\circ}$ and 3.9\,$^{\circ}$ for samples 2
and 3, respectively, as a consequence of bundling. The FWHM of the
twist distribution is found to be 2.8\,$^{\circ}$, 2.7\,$^{\circ}$,
and 3.1\,$^{\circ}$ for samples 1, 2, and 3, respectively.

The GISAXS measurements were performed at the beamline ID10 of the
European Synchrotron Radiation Facility (ESRF) using an X-ray energy
of 22~keV (wavelength $\lambda=0.5636\,\textrm{Å}).$ The incident
beam was directed at grazing incidence to the substrate. The chosen
grazing incidence angle was 0.2\,$^{\circ}$, i.\,e., about 2.5
times larger than the critical angle of the substrate, to avoid possible
complications of the scattering pattern typical for grazing incidence
X-ray scattering \cite{renaud09}. A two-dimensional detector Pilatus
300K (Dectris) placed at a distance of 2.38~m from the sample provided
a resolution of $8.06\times10^{-3}$~nm$^{-1}$.

%-----------------------------
\begin{figure}
\includegraphics[width=1\columnwidth]{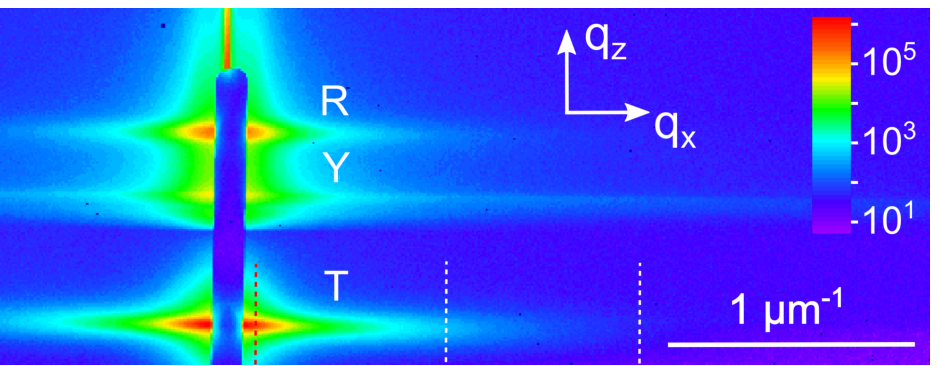}

\caption{GISAXS intensity from sample 1 as measured by a two-dimensional detector.
The scattering around the transmitted beam, the scattering around
the beam reflected from the substrate surface, and the Yoneda streak
are labeled as T, R, and Y, respectively. The vertical blue bar in
the middle of the scattering pattern is the beamstop. The three vertical
dotted lines mark the positions of the scans presented in Fig.~\ref{fig:columns}.
The color-coded scale bar represents the intensity in counts.}

\label{fig:MapDetector} 
\end{figure}

%----------------------------

Figure~\ref{fig:MapDetector} shows the GISAXS intensity measured
from sample~1. The scattering pattern comprises three horizontal
streaks. The small-angle scattering around the transmitted beam is
labeled as ``T'', while the scattering around the beam reflected
from the substrate surface is labeled as ``R''. Both streaks reveal
the same scattering intensity dependence on the lateral wavevector
$q_{x}$. The scattering around the transmitted beam possesses larger
intensity. For that reason, the T streaks is chosen here for the further
analysis. Besides the T and R streaks, the intensity distribution
in Fig.~\ref{fig:MapDetector} contains the Yoneda streak, marked
with ``Y'', which is located at the critical angle for total external
reflection. The chosen incidence angle allows us to separate well
the three different streaks, which facilitates the analysis of the
GISAXS intensity in the framework of kinematical scattering.

\section{Analysis of the measured intensities}

\label{sec:Analysis}

We use the specific features of the NWs as oriented long prisms to
improve the accuracy of the determination of the GISAXS intensity
$I(q_{x})$ from the measured maps. Since a single NW is a needle-like
object, its scattering intensity in the reciprocal space concentrates
in the plane perpendicular to the long axis of the NW. A random tilt
of a NW results in the respective tilt of the intensity plane. Hence,
one can expect that the spread of 4--5\,$^{\circ}$ in the directions
of the long axes of the NWs results in a sector of intensity in Fig.~\ref{fig:MapDetector}
with the width $\Delta q_{z}$ increasing proportional to $q_{x}$.

\begin{figure}
\includegraphics[width=0.8\columnwidth]{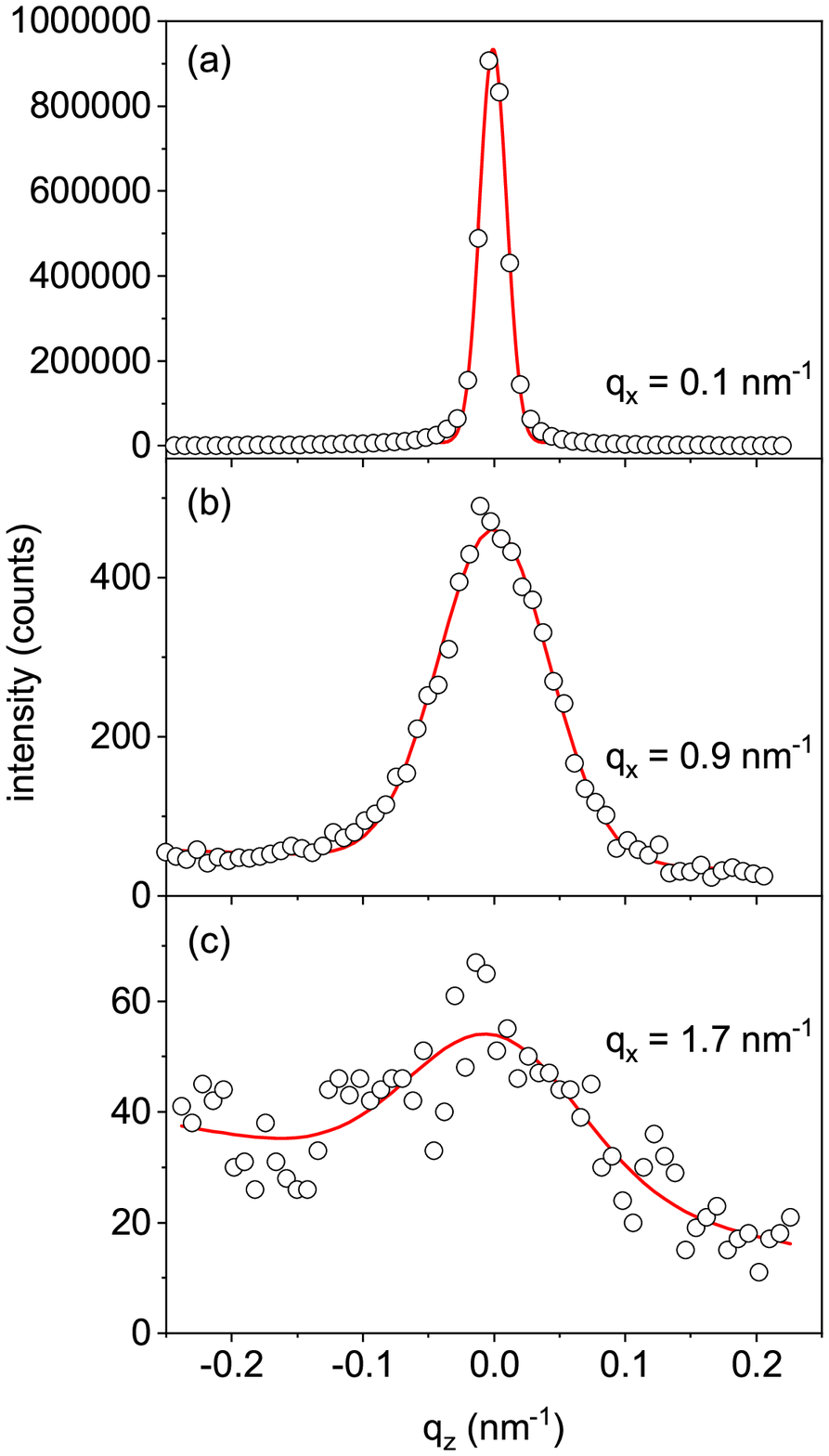}

\caption{Measured intensity profiles along the lines of constant $q_{x}$ values
marked by dotted lines in Fig.~\ref{fig:MapDetector} (circles) and
the respective Gaussian fits (lines).}

\label{fig:columns} 
\end{figure}

\begin{figure}
\includegraphics[width=0.9\columnwidth]{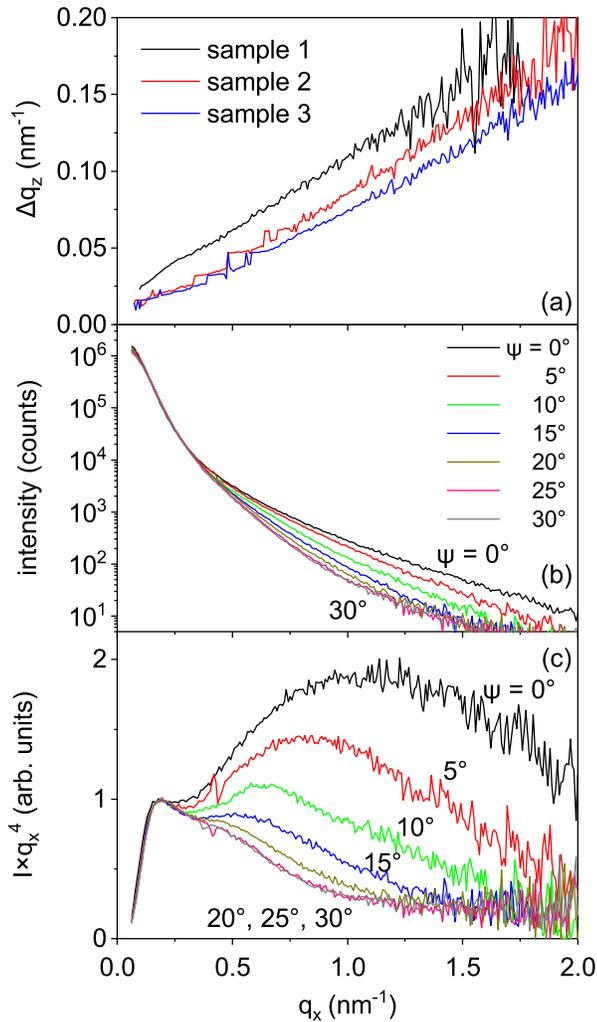}

\caption{(a) FWHMs of the intensity profiles $\Delta q_{z}$ as a function
of the wave vector $q_{x}$. (b) GISAXS intensity profiles of sample~1
as a function of the wave vector $q_{x}$ in dependence of the azimuthal
orientation $\psi$. (c) The same intensity profiles as in (b) but
plotted as $I(q_{x})q_{x}^{4}$ versus $q_{x}$.}

\label{fig:analysis} 
\end{figure}

Figure \ref{fig:columns} presents intensity profiles along the dotted
lines indicated in Fig.~\ref{fig:MapDetector}, i.e., scans at constant
values of $q_{x}$. These profiles are fitted by a Gaussian plus a
background that may linearly depend on $q_{z}$. The FWHMs of these
profiles $\Delta q_{z}$ are plotted in Fig.~\ref{fig:analysis}(a).
As expected, $\Delta q_{z}$ linearly increases with $q_{x}$. The
slopes $\Delta q_{z}/q_{x}$ give the angular ranges of the NW orientations
to be 5.9\,$^{\circ}$, 5.1\,$^{\circ}$, and 4.6\,$^{\circ}$
for samples 1, 2, and 3, respectively. These values are close to (albeit
somewhat larger) the widths of the orientational distributions measured
by Bragg diffraction, as described in Sec.~\ref{sec:Experiment}.

%------------------------------------------

%------------------------------------------

The fits in Fig.~\ref{fig:columns} help to improve the determination
of the scattering intensity $I(q_{x})$ both at small and large momenta
$q_{x}$. At small $q_{x}$, the intensity profiles are narrow and
the peak intensity has to be determined from just a few data points.
At large $q_{x}$, the intensity is low and the background is comparable
to the signal. After performing the fits of the cross-sectional profiles
(i.e., along the $q_{z}$ direction) shown in Fig.~\ref{fig:columns}
and establishing the linear dependence of the FWHM $\Delta q_{z}$
on $q_{x}$, we make one more step to improve the accuracy. Linear
fits are made for the $\Delta q_{z}$ on $q_{x}$ dependencies plotted
in Fig.~\ref{fig:analysis}(a). Then, the fits of the $q_{z}$ profiles
shown in Fig.~\ref{fig:columns} are repeated, now with the FWHMs
fixed at the values obtained from the linear fits. In this way, the
number of free parameters in the Gaussian fits is decreased, and the
intensity $I(q_{x})$ is determined more accurately. This intensity
is used in the further analysis.

%-----------------------------------------------------------

%-----------------------------------------------------------

Figure~\ref{fig:analysis}(b) presents the GISAXS intensity $I(q_{x})$
measured on sample 1 in dependence on its azimuthal orientation $\psi$.
The sample orientation $\psi=0$ corresponds to the incident X-ray
beam along a GaN$\left\langle 11\bar{2}0\right\rangle $ direction,
so that the scattering vector (the $x$-axis direction) is along $\left\langle 1\bar{1}00\right\rangle $,
which is the normal to the NW facets. Figure \ref{fig:analysis}(b)
comprises the measurements obtained on the rotation of sample 1 about
the normal to the substrate surface (i.e., about the direction of
the long axes of the NWs) from $\psi=0\,^{\circ}$ to 30\,$^{\circ}$
with a step of 5\,$^{\circ}$. Since the sample has a rectangular
shape and the illuminated area varies on rotation, the curves are
scaled to obtain the same intensity in the small-$q_{x}$ range. The
scaling factors differ less than by a factor of 2. The azimuthal dependence
of the intensity at large $q_{x}$ is evident from the plot.

In the case of the reflected beam (the streak ``R'' in Fig.~\ref{fig:MapDetector}),
an identical analysis of the intensity (not shown here) results in
curves close to those shown in Fig.~\ref{fig:analysis}(b). Thus,
we observe the same azimuthal dependence of the intensity but with
a smaller total intensity and a higher level of noise. Because of
this reason, for the further analysis presented in the paper, we exclusively
consider the intensity distributions around the transmitted beam.

Since we expect Porod's law $I(q_{x})\sim q_{x}^{-4}$ to be satisfied
at large $q_{x}$, we plotted in Fig.~\ref{fig:analysis}(c) the
same data as $I(q_{x})q_{x}^{4}$ versus $q_{x}$, which would tend
to a constant value for large $q_{x}$. Surprisingly, a strong deviation
from Porod's law is observed. Furthermore, the data do not only deviate
from Porod's law, but also exhibit a strong azimuthal dependence.
In other to explain this unexpected behavior, in the next section,
we develop a Monte Carlo method to calculate the scattering intensity.

\section{Calculation of the scattering intensity}

\label{sec:Calculation}

\subsection{Scattering amplitude of a prism}

\label{subsec:Scattering-prism}

We calculate first the scattering amplitude (form factor) of a NW
$A(\mathbf{q})$ in a coordinate system linked to the NW, i.e., with
$z$-axis in the direction of the long axis of the NW. Hence, the
cross-section of the NW is in the $xy$ plane. Next, we will consider
in Sec.~\ref{subsec:3Ddistribution} a transformation of the wave
vectors from the laboratory frame to the NW coordinate system, and
perform an average of the intensities $\left|A(\mathbf{q})\right|^{2}$
over different NW orientations.

The scattering amplitude of a NW is given by its form factor 
\begin{equation}
A(\mathbf{q})=\int_{V}\exp(i\mathbf{q}\cdot\mathbf{r})\,\mathrm{d}\mathbf{r},\label{eq:1}
\end{equation}
where the integral is calculated over the NW volume $V$. Since the
NW is a prism, the scattering amplitude can be represented as a product
of the components along the NW axis and in the plane perpendicular
to it, $A(\mathbf{q})=A_{\parallel}(q_{\parallel})A_{\perp}(\mathbf{q_{\perp}})$.
The longitudinal component is simply 
\begin{equation}
A_{\parallel}(q_{\parallel})=\mathrm{sinc}(q_{\parallel}L/2),\label{eq:2}
\end{equation}
where $\mathrm{sinc}(x)=(\sin x)/x$ and $L$ is the NW length.

%-----------------------------------
\begin{figure}
\includegraphics[width=1\columnwidth]{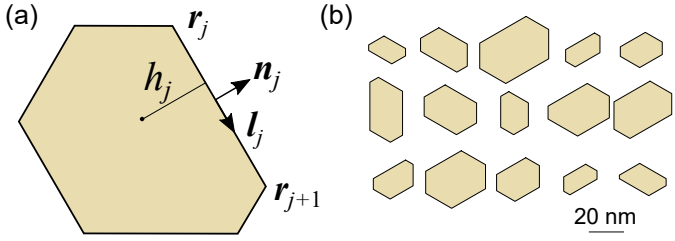}

\caption{(a) A hexagon with vertices $\mathbf{r}_{j}$ and the unit vectors
along and normal to the side $\mathbf{l}_{j}$ and $\mathbf{n}_{j}$.
The distance from the hexagon center to its side is $h_{j}$. (b)
Examples of randomly generated hexagons used to simulate the scattering
from sample~1.}

\label{fig:polygon} 
\end{figure}

%-----------------------------------

The calculation of the transverse component $A_{\perp}(\mathbf{q_{\perp}})$
can be reduced to a sum over the vertices, as it was initially shown
for faceted crystals by \citeasnoun{laue36} and used nowadays to
calculate form factors of nanoparticles \cite{vartanyants08,renaud09,pospelov20}.
Specifically, \citeasnoun{laue36} proposed to reduce, using Gauss'
theorem, the volume integral (\ref{eq:1}) to the integrals over the
facets; application of Gauss' theorem to these area integrals reduces
them to integrals over the edges, which, in turn, can be taken by
parts and expressed through the coordinates of the vertices.

For a planar polygon, the form factor reads 
\begin{equation}
A_{\perp}(\mathbf{q_{\perp}})=\frac{1}{q_{\perp}^{2}}\sum_{j}\frac{\mathbf{q}_{\perp}\cdot\mathbf{n}_{j}}{\mathbf{q}_{\perp}\cdot\mathbf{l}_{j}}\left(e^{i\mathbf{q}_{\perp}\cdot\mathbf{r}_{j+1}}-e^{i\mathbf{q}_{\perp}\cdot\mathbf{r}_{j}}\right),\label{eq:3}
\end{equation}
where the sum runs over the vertices and, as illustrated in Fig.~\ref{fig:polygon}(a),
$\mathbf{r}_{j}$ are coordinates of the vertices, $\mathbf{l}_{j}$
and $\mathbf{n}_{j}$ are unit vectors along the polygon side between
the vertices $\mathbf{r}_{j}$ and $\mathbf{r}_{j+1}$ and normal
to it, respectively. \citeasnoun{lee83} proposed another expression
for the form factor, 
\begin{equation}
A_{\perp}(\mathbf{q_{\perp}})=\sum_{j}e^{i\mathbf{q}_{\perp}\cdot\mathbf{r}_{j}}\frac{\left(\mathbf{l}_{j}\times\mathbf{l}_{j-1}\right)\cdot\mathbf{N}}{\left(\mathbf{q}_{\perp}\cdot\mathbf{l}_{j}\right)\left(\mathbf{q}_{\perp}\cdot\mathbf{l}_{j-1}\right)},\label{eq:4}
\end{equation}
where $\mathbf{N}$ is the unit vector normal to the polygon plane,
and \citeasnoun{wuttke17} explicitly showed the identity of the
expressions (\ref{eq:3}) and (\ref{eq:4}). Equation (\ref{eq:3})
makes it possible to easily resolve the numerical uncertainty $0/0$
that arises at $\mathbf{q}_{\perp}\cdot\mathbf{l}_{j}=0$. Since $\mathbf{l}_{j}=\left(\mathbf{r}_{j+1}-\mathbf{r}_{j}\right)/\left|\mathbf{r}_{j+1}-\mathbf{r}_{j}\right|$,
we have in the limit $\mathbf{q}_{\perp}\cdot\mathbf{l}_{j}\rightarrow0$
\begin{equation}
\frac{1}{\mathbf{q}_{\perp}\cdot\mathbf{l}_{j}}\left(e^{i\mathbf{q}_{\perp}\cdot\mathbf{r}_{j+1}}-e^{i\mathbf{q}_{\perp}\cdot\mathbf{r}_{j}}\right)\rightarrow ie^{i\mathbf{q}_{\perp}\cdot\mathbf{r}_{j}}\left|\mathbf{r}_{j+1}-\mathbf{r}_{j}\right|.\label{eq:4a}
\end{equation}

\begin{figure}
\includegraphics[width=1\columnwidth]{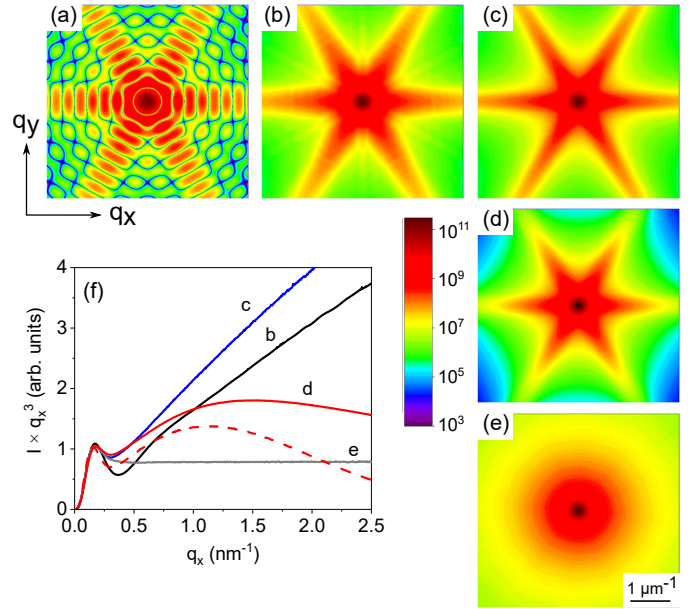}

\caption{Scattering intensity from (a) a regular hexagon with a side length
of 12~nm, (b) a distribution of regular hexagons with the average
side length of 12~nm and a standard deviation of the side lengths
of 4~nm, (c) a distribution of distorted hexagons as shown in Fig.~\ref{fig:polygon}(b),
(d) the same distribution as in (c) but with a side facet roughness
$\sigma=0.6$~nm added according to Eq.~(\ref{eq:5a}), (e) a distribution
of randomly oriented distorted hexagons. The color-coded scale bar
representing the intensity is applicable to (a)--(e). (f) Radial
intensity distributions from (b)--(e) in the directions of the intensity
maxima, the product $Iq^{3}$ is plotted. The curves are labeled by
the same symbols as the respective maps. The effect of roughness is
illustrated in (f) by two curves: the full red curve corresponds to
the geometric distribution of the atomic steps on the side facets
and the dashed red curve to the Poisson distribution, both possessing
the same rms roughness of $\sigma=0.6$~nm.}

\label{fig:stars} 
\end{figure}

Figure~\ref{fig:stars}(a) shows the intensity distribution calculated
by Eq.~(\ref{eq:3}) for a regular hexagon with a side length of
12~nm. The intensity is higher in the directions of the side normals
and oscillates due to interference from opposite sides of the hexagon.
Figure~\ref{fig:stars}(b) shows a Monte Carlo calculation of the
average intensity from hexagons of different sizes. A lognormal distribution
of the lengths of the hexagon sides is taken with the same mean value
of\textit{\emph{ 12~nm}} and a standard deviation of 4~nm.

%-------------------------------------

%-------------------------------------

The radial intensity distribution in the direction along the intensity
maximum is presented in Fig.~\ref{fig:stars}(f) by the black line.
The intensity distribution is presented as the product $Iq_{x}^{3}$,
which would be constant at large $q_{x}$ for an ensemble of randomly
oriented hexagons (as well as for other two-dimensional objects with
rigid boundaries) after averaging over all possible orientations.
As stated above, the intensity maxima in Fig.~\ref{fig:stars}(b)
correspond to the directions normal to the sides of the hexagon. They
possess, at large $q_{x}$, a $I\propto q_{x}^{-2}$ dependence due
to a steplike variation of the density at a planar surface. Hence,
in Fig.~\ref{fig:stars}(f), we observe a linear increase of the
intensity for large values of $q_{x}$ (black line).

The local maximum at $q_{m}=0.175$~nm$^{-1}$ in Fig.~\ref{fig:stars}(f)
is related to the mean length of the side facet of the hexagons $a=12$~nm
as $a\approx2.1/q_{m}$, which allows us to determine the hexagon
size directly from the plots of $Iq_{x}^{3}$ versus $q_{x}$ (in
the three-dimensional case of hexagonal prisms, the same formula is
applicable for the maximum in $Iq_{x}^{4}$ versus $q_{x}$ plot,
see Sec.~\ref{subsec:3Ddistribution}). For comparison, the form
factor of a circle of radius $R$ gives $I(q)q^{3}\propto J_{1}^{2}(qR)$,
where $J_{1}(x)$ is the Bessel function. The first maximum of $J_{1}(x)$
at $x\approx1.84$ gives the circle radius $R\approx1.84/q_{m}$.
We can relate a hexagon and a circle even closer, by defining an effective
radius $R_{a}$ of a circle possessing the same area as the hexagon
with a side length $a$. Then, we have $R_{a}=(3\sqrt{3}/2\pi)^{1/2}a\approx0.91a$
and $R_{a}\approx1.9/q_{m}$, with the proportionality coefficient
very close to the case of a circle.

With the form factor defined by the positions of the vertices according
to either Eq.~(\ref{eq:3}) or Eq.~(\ref{eq:4}), we are not restricted
to regular hexagons but can take into account the real cross-sectional
shapes of the NWs. Since the side facets of the NWs are GaN\{$1\bar{1}00$\}
planes making an angle of 60\,$^{\circ}$ to each other, we build
the hexagons as shown in Fig.~\ref{fig:polygon}(a): random heights
$h_{j}$ are taken in the directions normal to the facets. Then, we
check that the generated hexagon is convex, and discard it otherwise.
Figure \ref{fig:polygon}(b) presents examples of randomly generated
hexagons with the same orientation of their sides. The distribution
of the hexagon shapes is chosen to simulate sample 1 and further described
in Sec.~\ref{sec:Results}. The intensity map obtained from this
distribution of hexagons is shown in Fig.~\ref{fig:stars}(c). The
respective radial intensity distribution is presented in Fig.~\ref{fig:stars}(f)
by the blue line. One can see that the black and the blue lines in
Fig.~\ref{fig:stars}(f) are remarkably different. In particular,
the hexagon shape distribution notably reduces the dip in the intensity.
Therefore, the distortion of the hexagons can be deduced from the
intensity plots.

\subsection{Roughness of the side facets}

The side facets of GaN NWs are atomically flat \cite{stoica08,ristic08}
but may have atomic steps. The radial growth of these NWs presumably
proceeds by step flow, with the motion of steps from the NW top, where
they are nucleated, down along the side facets \cite{garrido13}.
Random steps across the side facets can be treated as facet roughness
in the same way as it is done in the calculation of crystal truncation
rods \cite{Robinson1986a,robinson92}.

A step of height $d_{0}$ shifts the $j$th side of the polygon in
Fig.~\ref{fig:polygon} by a vector $d_{0}\mathbf{n}_{j}$ in the
direction of the facet normal. Hence, the $j$th term in the sum (\ref{eq:3})
acquires an additional factor $\exp\left(id_{0}\mathbf{q}_{\perp}\cdot\mathbf{n}_{j}\right)$.
Random steps give rise to a factor $R_{j}=R(\mathbf{q}_{\perp}\cdot\mathbf{n}_{j})$,
where the function $R(q)$ is defined as 
\begin{equation}
R(q)=\sum_{m=0}^{\infty}p_{m}\exp\left(imqd_{0}\right),\label{eq:5}
\end{equation}
here $p_{m}$ are the probabilities of the shift of the side facet
by $m$ steps. Hence, the function $R(q)$ is the characteristic function
of the probabilities $p_{m}$.

Consider the geometric probability distribution $p_{m}=(1-\beta)\beta^{m}$
with the parameter $\beta<1$. It describes a flat surface with a
fraction $\beta$ one step higher, its fraction $\beta$ is, in turn,
one step higher, and so on \cite{Robinson1986a}. The root mean squared
(rms) roughness is $\sigma=d_{0}\sqrt{\beta}/(1-\beta)$ and the corresponding
characteristic function is 
\begin{equation}
R(q)=\frac{1-\beta}{1-\beta\exp(iqd_{0})}.\label{eq:5a}
\end{equation}

The Poisson probability distribution $p_{m}=\exp(-\mu)\mu^{m}/m!$
gives rise to the rms roughness $\sigma=d_{0}\sqrt{\mu}$ and the
characteristic function is 
\begin{equation}
R(q)=\exp\left[-\mu\left(1-e^{iqd_{0}}\right)\right].\label{eq:5b}
\end{equation}

We stress here that the $j$th term in the sum (\ref{eq:3}) is multiplied
with a complex factor $R_{j}=R(\mathbf{q}_{\perp}\cdot\mathbf{n}_{j})$
that depends on the orientation of the respective facet. This is different
from a common treatment of the surface roughness, which involves a
single factor $\left|R\right|^{2}$. Particularly, the Poisson probability
distribution gives for $qd_{0}\ll1$ the factor $\left|R\right|^{2}=\exp\left(-\sigma^{2}q^{2}\right)$.
\citeasnoun{buttard13} used such a factor to describe the effect
of the roughness on the scattered intensity from Si NWs, by analogy
to the roughness of planar surfaces, and arrived at an rms roughness
$\sigma$ of 1~nm for their samples. We use Eq.~(\ref{eq:3}) in
further calculations with the complex factors $R_{j}$ in each term
of the sum.

Figure~\ref{fig:stars}(d) shows the scattering intensity distribution
obtained with the roughness factors given by Eq.~(\ref{eq:5a}).
The rms roughness is taken to be $\sigma=0.6$~nm and the step height
$d_{0}$ is that of the atomic steps on the GaN($1\bar{1}00$) facet,
$d_{0}=a_{0}\sqrt{3}/2$, where $a_{0}=0.319$~nm is the GaN lattice
spacing. Strictly speaking, the roughness factors given by Eq.~(\ref{eq:5a})
or Eq.~(\ref{eq:5b}) are derived for a prism which has in each cross-section
a hexagon with straight sides. It describes a variation of the cross-section
of the prism along its length and does not make sense for two-dimensional
objects. Hence, the intensity distribution shown in Fig.~\ref{fig:stars}(d)
corresponds to the prisms with perfectly aligned long axes.

The solid red line in Fig.~\ref{fig:stars}(f) shows the radial intensity
distribution obtained from the map shown in Fig.~\ref{fig:stars}(d)
in the direction of maximum intensity, calculated using the roughness
factors for the geometric probability distribution given by Eq.~(\ref{eq:5a}).
The dashed red line in Fig.~\ref{fig:stars}(f) shows the intensity
from the same distribution of hexagons but calculated using the roughness
factors derived from the Poisson probability distribution, Eq.~(\ref{eq:5b}).
The rms roughness is taken the same in both cases, $\sigma=0.6$~nm.
One can see that the roughness qualitatively changes the intensity
at $q\sigma>1$. Hence, the rms roughness $\sigma$ can be obtained
from the intensity plots. Moreover, the intensity curves are fairly
sensitive to the choice of the probability distribution. The crystal
truncation rods from planar surfaces possess a similar sensitivity
to the choice of the roughness model \cite{walko00}. Our modeling
of the scattering from GaN NWs presented in Sec.~\ref{sec:Results}
shows that the geometric probability distribution provides a better
agreement with the experimental data.

\subsection{Orientational distribution of the NWs}

\label{subsec:3Ddistribution}

The scattering intensity is measured as a function of the wave vector
$\mathbf{q}$ in the laboratory frame (see Fig.~\ref{fig:MapDetector}).
We need to find the components $(q_{\parallel},\mathbf{q}_{\perp})$
of this vector in the frame given by the long axis of the NW and the
normal to one of its side facets. Let us consider first the simple
case of the two-dimensional rotation of the hexagons (or perfectly
aligned prisms in the plane normal to their long axes). The unit vector
normal to a hexagon side (or the prism facet) can be written as 
\begin{equation}
\mathbf{n}=(\cos\psi,\sin\psi,0),\label{eq:7}
\end{equation}
where $\psi$ is a polar angle (defined modulo 60\,$^{\circ}$) with
respect to a reference orientation. The unit vector along the hexagon
side is, respectively, $\mathbf{l}=(-\sin\psi,\cos\psi,0)$. The components
$q_{n}$, $q_{l}$ of the two-dimensional vector $\mathbf{q}_{\perp}$
are determined simply as $q_{n}=\mathbf{q_{\perp}}\cdot\mathbf{n}$
and $q_{l}=\mathbf{q_{\perp}}\cdot\mathbf{l}$.

Figure~\ref{fig:stars}(e) presents a Monte Carlo calculation of
the intensity for the distribution of distorted hexagons described
above and sketched in Fig.~\ref{fig:polygon}(b), after an average
over the orientations $\psi$ uniformly distributed from 0\,$^{\circ}$
to 360\,$^{\circ}$. The corresponding radial intensity distribution,
shown in Fig.~\ref{fig:stars}(f) by a gray line, follows the two-dimensional
Porod's law $I(q)\propto q^{-3}$ at large $q$. At small $q$, it
coincides with the intensity distribution for the oriented hexagons.

For the three-dimensional distribution of the NW orientations, inherent
to the spontaneous formation of GaN NWs on Si(111) (see Fig.~\ref{fig:SEMimages}),
we define a unit vector in the direction of the long NW axis as 
\begin{equation}
\mathbf{e}_{\parallel}=(\sin\theta\cos\phi,\sin\theta\sin\phi,\cos\theta),\label{eq:8}
\end{equation}
where $\phi$ and $\theta$ are the azimuthal angle of tilt and its
polar angle, respectively. The unit vector normal to the facet is
defined to be orthogonal to $\mathbf{\mathbf{e}_{\parallel}}$ and
possessing the same projection on the horizontal plane as in Eq.~(\ref{eq:7}):
\begin{equation}
\mathbf{n}=(\xi\cos\psi,\xi\sin\psi,s\sqrt{1-\xi^{2}}),\label{eq:9}
\end{equation}
where $\xi=\left[1+\tan^{2}\theta\cos^{2}\left(\phi-\psi\right)\right]^{-1/2}$
and $s$ is the sign of $-\cos\left(\phi-\psi\right)$. Since the
tilt angle $\theta$ does not exceed a few degrees, the difference
between Eqs.~(\ref{eq:7}) and (\ref{eq:9}) is unessential. The
vector $\mathbf{l}$ is defined as a vector product $\mathbf{l}=\mathbf{e}_{\parallel}\times\mathbf{n}$.

We calculate the scattering intensity by the Monte Carlo method. It
enables a simultaneous integration over the distributions of the NW
lengths, their cross-sectional sizes and shapes, and the orientations
of the NW long axis as well as those of their side facets. The calculations
take fairly little time. It takes less than a minute on a single CPU
core of a standard PC to calculate an intensity curve with the accuracy
sufficient to make estimates. The smooth curves presented in the paper
took less than an hour of CPU time each.

We take the mean NW lengths obtained from the scanning electron micrographs
and given in Sec.~\ref{sec:Experiment}. A large scattering in the
NW lengths is evident from Fig.~\ref{fig:SEMimages}. The length
distribution is assumed to be lognormal with a standard deviation
of 20\% from the respective average lengths. For the facet orientation
angle $\psi$, we take a normal distribution with the FWHM determined
by the in-plane X-ray diffraction scans (see Sec.~\ref{sec:Experiment}).
The average value of $\psi$ is given by the orientation of the incident
X-ray beam with respect to the NW facets {[}see Figs.~\ref{fig:analysis}(b)
and \ref{fig:analysis}(c){]}.

The integration over the orientations of the long axis of the NWs
is an integration over a solid angle, i.e., the integral of the intensity
from a single NW with $P(\theta)\sin\theta\,\mathrm{d}\theta\mathrm{d}\phi$,
where $P(\theta)$ is the probability density distribution of the
tilt angle $\theta$. The azimuthal angle $\phi$ is uniformly distributed
from 0 to $2\pi$. We take a normal distribution of the tilt angles,
$P(\theta)=\sqrt{2/\pi}(\Delta\theta)^{-1}\exp[-\theta^{2}/2(\Delta\theta)^{2}]$.
The standard deviation $\Delta\theta$ can be obtained from the slopes
of the straight lines in Fig.~\ref{fig:analysis}(a) or from the
FWHMs of the symmetric Bragg reflections GaN 0002, as discussed in
Sec.~\ref{sec:Analysis}. The widths $\Delta q_{z}$ in Fig.~\ref{fig:analysis}(a)
and the FWHMs of the Bragg reflections take into account the tilts
in all directions, so that $\Delta\theta$ is obtained from the half
width at half maximum (HWHM) and varies from 2.5\,$^{\circ}$ for
sample 1 to 2\,$^{\circ}$ for sample 3. Since the tilt angle $\theta$
is small, we can take $\sin\theta\approx\theta$ and proceed to an
integral over a new variable $y=\theta^{2}$. Then, the integral is
taken with $\tilde{P}(y)\,\mathrm{d}y\mathrm{d}\phi$, where $\tilde{P}(y)\propto\exp[-y/2(\Delta\theta)^{2}]$.
Hence, we generate $y$ as an exponentially distributed random number
with the unit dispersion and calculate $\theta=\sqrt{2y}\Delta\theta.$

If the NW orientations are completely random, i.e., the angles $\phi$
and $\psi$ vary from 0 to $2\pi$ and the angle $\theta$ from 0
to $\pi$ uniformly and independently, the small-angle scattering
intensity at $q\gg2\pi/a$, where $a$ is the width of the side facet,
follows Porod's law $I(q)\propto q^{-4}$. However, since the NWs
are long prisms, the scattering intensity from a single NW of length
$L$ with its long axis in $z$-direction concentrates in the reciprocal
space in a disk of the width $\Delta q_{z}=2\pi/L$. We have seen
in Sec.~\ref{sec:Analysis} that the scattering from the oriented
NWs is limited by $\Delta q_{z}/q_{x}<\Delta\theta$, where $\Delta\theta$
is the angular range of orientations. As long as $2\pi/(Lq_{x})<\Delta\theta$,
the oriented NWs give the same intensity in the $x$-direction as
fully randomly oriented ones. Therefore, Porod's law is satisfied
for $q_{x}>2\pi/(L\Delta\theta)$.

%------------------------------------------------------
\begin{figure}
\includegraphics[width=0.8\columnwidth]{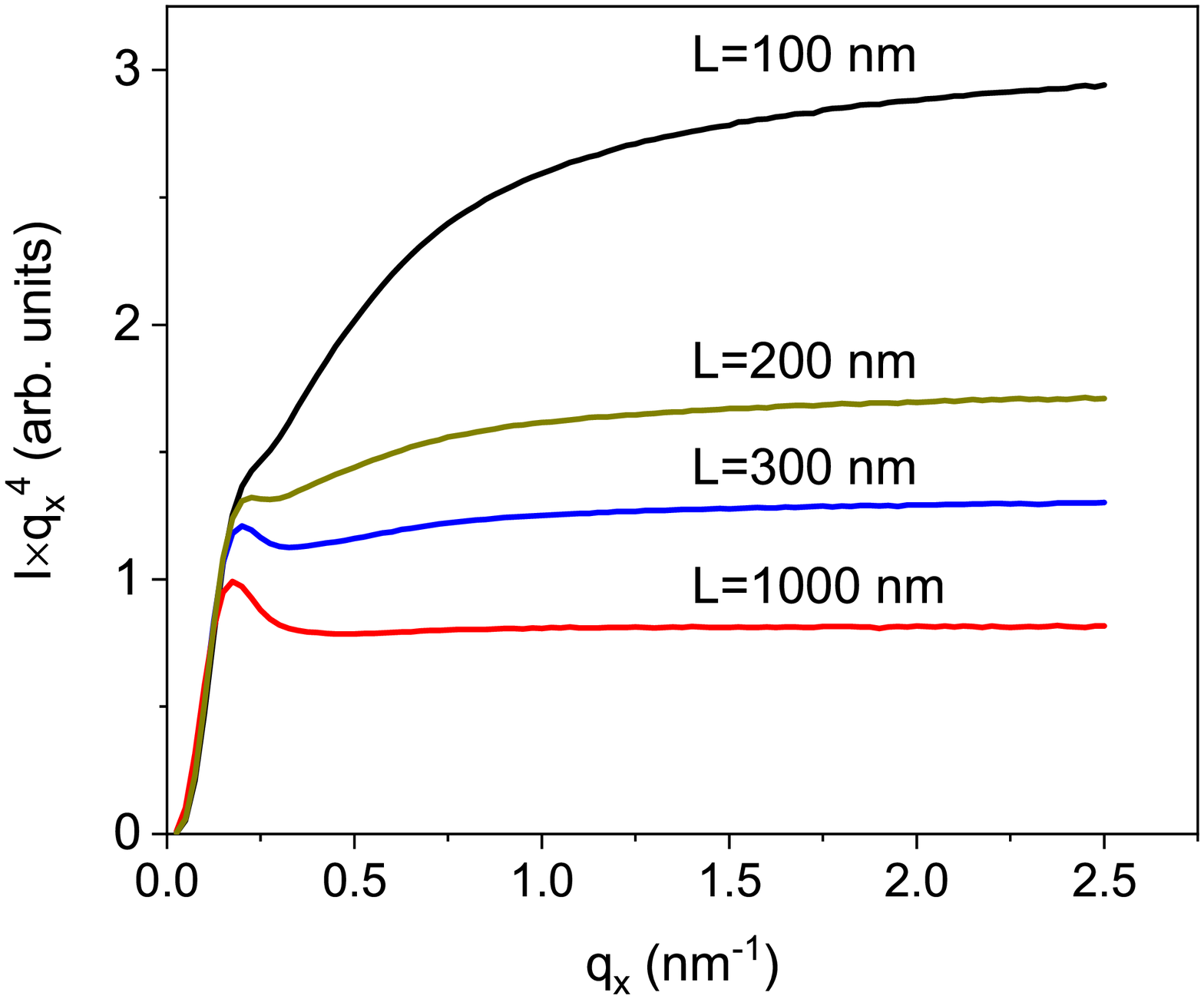}

\caption{Monte Carlo calculation of the small-angle scattering intensity from
an ensemble of NWs with a 5\,$^{\circ}$ wide range of orientations
of the long axes and random orientation of the side facets. The width
of the side facets is 12~nm and NW lengths vary from 100 to 1000~nm.}
\label{fig:PorodLaw3D} 
\end{figure}

%------------------------------------------------------

Figure \ref{fig:PorodLaw3D} presents the Monte Carlo calculation
of the small-angle scattering intensity from NWs of different lengths
and the same width of the tilt angle distribution $\Delta\theta=2.5\,^{\circ}$
corresponding to that of sample 1. Particularly, for the NWs with
a length $L=200$~nm, the condition derived above reads $q_{x}>0.7$~nm$^{-1}$,
which is in a good agreement with the region of constant $I(q_{x})q_{x}^{4}$
in Fig.~\ref{fig:PorodLaw3D}. Hence, the limited range of the tilt
angles in the NW ensemble does not prevent reaching Porod's law even
for the relatively short NWs of sample 1. We remind that the curves
in Fig.~\ref{fig:PorodLaw3D} are calculated by averaging over the
tilt azimuth $\phi$ and the facet orientation $\psi$ varying from
0 to $2\pi$. Further Monte Carlo calculations, taking into account
the orientational ordering of the side facets of the epitaxially grown
GaN NWs, are presented in the next section.

\section{Results}

\label{sec:Results}

\begin{figure}
\includegraphics[width=1\textwidth]{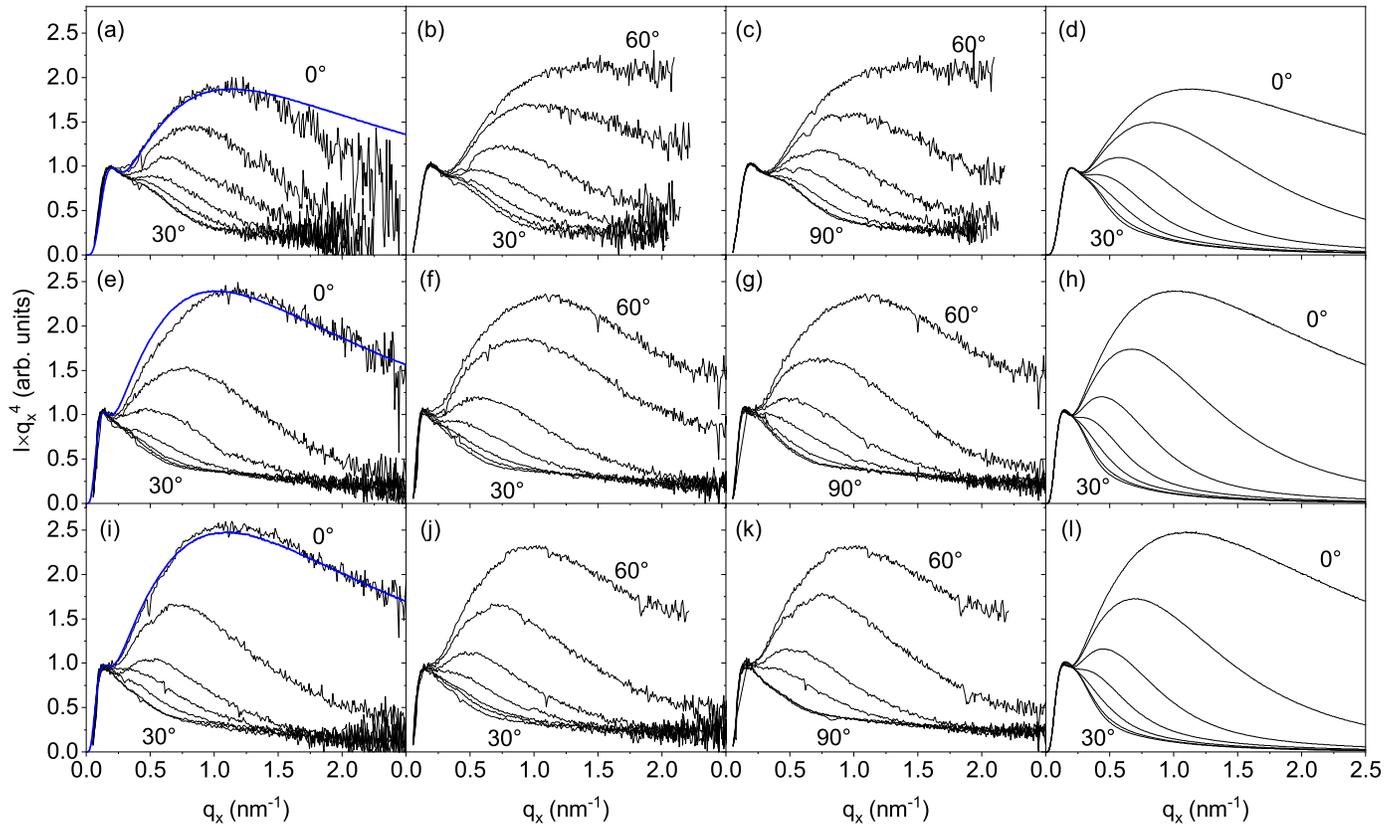}

\caption{The measured GISAXS intensity (three left columns) and Monte Carlo
simulations (right column) for samples 1 (a--d), 2 (e--h), and 3
(i--l). The measurements are performed for different mean orientation
angles $\psi$ of the side facets of the NWs with respect to the X-ray
beam, namely, from 0\,$^{\circ}$ to 90\,$^{\circ}$ with steps
of 5\,$^{\circ}$. For clarity, these measurements are presented
in three different panels (from 0\,$^{\circ}$ to 30\,$^{\circ}$,
from 30\,$^{\circ}$ to 60\,$^{\circ}$, and from 60\,$^{\circ}$
to 90\,$^{\circ}$). The intensities are plotted as $I(q_{x})q_{x}^{4}$
versus $q_{x}$ to highlight deviations from Porod's law. The curves
calculated at $\psi=0^{\circ}$ for each sample are repeated as blue
curves in the left column, for a direct comparison of the calculated
and the measured curves.}

\label{fig:Iq4} 
\end{figure}

Figure \ref{fig:Iq4} presents the results of the systematic GISAXS
measurements on samples 1--3. The measurements and their analysis
are described in the sections \ref{sec:Experiment} and \ref{sec:Analysis},
respectively. The samples have been measured with the azimuthal orientation
$\psi$ varying from 0\,$^{\circ}$ to 90\,$^{\circ}$ with steps
of 5\,$^{\circ}$. Each measurement provided a map similar to the
one in Fig.~\ref{fig:MapDetector}, and the intensity $\mathcal{I}(q_{x},q_{z})$
around the transmitted beam has been analyzed by fitting every scan
of a constant $q_{z}$ by a Gaussian, as shown in Fig.~\ref{fig:columns}.
The peak values of the $q_{z}$ scans obtained in this fit provided
the intensity $I(q_{x})$. It is presented in Fig.~\ref{fig:Iq4}
as the product $I(q_{x})q_{x}^{4}$ versus $q_{x}$, to reveal deviations
from Porod's law.

%------------------------
\onecolumn

\twocolumn %------------------------

Since GaN NWs grow epitaxially on Si(111), the ensemble possesses
a 6-fold orientational symmetry. A systematic variation of the intensity
curves depending on the azimuthal sample orientation $\psi$ is evident
from Fig.~\ref{fig:Iq4}: within the statistical error of the measurements,
the orientations $\psi$ and $\psi+60\,^{\circ}$, as well as $\pm\psi$
or $30\,^{\circ}\pm\psi$, are equivalent. Hence, the Monte Carlo
modeling presented in the right column of Fig.~\ref{fig:Iq4} has
been performed for the angle $\psi$ from 0\,$^{\circ}$ to 30\,$^{\circ}$
with the same step of 5\,$^{\circ}$.

In the Monte Carlo calculations, we take the values of the NW lengths
$L$, the range of the tilt angles $\theta$, and the range of the
side facet orientations $\psi$ at the values experimentally determined
in Secs.~\ref{sec:Experiment} and \ref{sec:Analysis}. The parameters
of the NW ensemble to be determined from the modeling are the mean
width of the side facets $a$, its variation as well as the variation
of the cross-sectional shapes of the NWs, and the roughness of the
side facets. We have seen in Sec.~\ref{sec:Analysis} that these
parameters affect the calculated curves in qualitatively different
ways. The mean facet size $a$ determines the position of the local
maximum of the curves at $q_{x}\approx0.17$~nm$^{-1}$, which corresponds
to a side facet width of about 12~nm. The depth of the dip between
this maximum and the rise of the curves at larger $q_{x}$ is controlled
by the width of the facet size distribution and the shape distribution
of the cross-sections. The decrease of $Iq_{x}^{4}$ at large $q_{x}$
is caused by the roughness of the side facets.

The distorted cross-sections of the NWs are modeled in the Monte Carlo
study as described in Sec.~\ref{subsec:Scattering-prism}. The heights
$h_{j}$ shown in Fig.~\ref{fig:polygon}(a) are generated on random
around a mean value. Figure \ref{fig:polygon}(b) exemplifies the
shapes of the NWs used in the simulation of sample 1. The right column
in Fig.~\ref{fig:Iq4} presents the Monte Carlo calculation of the
small-angle scattering intensity for samples 1--3. For a direct comparison
of the calculated and the measured intensities, the curves calculated
for each sample at $\psi=0{}^{\circ}$ are repeated as blue lines
in the left column of the figure.

\begin{figure}
\includegraphics[width=1\columnwidth]{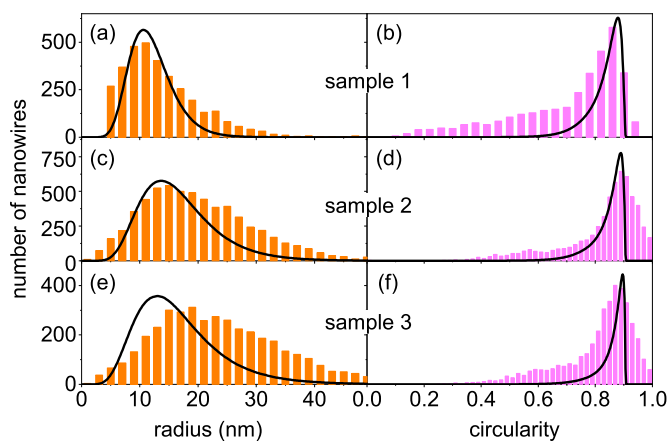}

\caption{Distributions of the NW radii and circularities of samples 1--3 obtained
from the analysis of the top-view scanning electron micrographs (histograms)
and the Monte Carlo modeling (lines).}

%------------------------------------------------------------------

\label{fig:histograms} 
\end{figure}

For each generated NW, we calculate the cross-sectional area $A$
and the perimeter $P$. Then, we determine out of these parameters
the radius $R$ from $A=\pi R^{2}$ and the circularity $C=4\pi A/P^{2}$.
The circularity thus defined is $C=1$ for a circle, $C=\pi\sqrt{3}/6\approx0.907$
for a regular hexagon, and $C\ll1$ for irregular shapes. These parameters,
radius and circularity, can be obtained from scanning electron micrographs
and are objective NW shape descriptors as discussed elsewhere \cite{brandt14}.
The lines in Fig.~\ref{fig:histograms} show the distributions of
the radius and the circularity obtained in the simulations.

The distributions of the cross-sectional radii and circularities of
the NW ensembles have also been interdependently obtained by analyzing
top-view scanning electron micrographs similar to those shown in Fig.~\ref{fig:SEMimages}(d--f).
The analysis has been performed using the open-source software ImageJ
\cite{schneider12}, as described in detail by \citeasnoun{kaganer16bundling}
in their Supporting Information. The distribution of the radius obtained
from the modeling of the GISAXS intensity for sample 1 in Fig.~\ref{fig:histograms}(a)
is fairly close to the distribution derived from the micrographs.
The circularity distribution obtained from the micrographs is, however,
extended towards smaller values indicating a higher density of NWs
with elongated cross-sectional shapes. Such a discrepancy can be attributed
to an artifact caused by the NW tilt. Specifically, the scanning electron
micrographs exhibit a very little difference in brightness between
the top facet and the top part of the side facet of the NW, so that
ImageJ treats both regions together, i.\,e., as extended intensity
spots.

In contrast to sample 1, the NW radii obtained from the Monte Carlo
simulations of the scattering intensity from samples 2 and 3, see
Figs.~\ref{fig:histograms}(c) and \ref{fig:histograms}(e), are
smaller than those derived from the analysis of the scanning electron
micrographs, and the discrepancy increases with increasing NW length.
We remind that the mean NW radius can be directly derived from the
position $q_{x}$ of the local maximum in the experimental curves
presented in Fig.~\ref{fig:Iq4}. It remains at about $q_{x}\approx0.17$~nm$^{-1}$
and only slightly shifts to smaller values (and hence to larger radii)
as the NW length increases from sample 1 to sample 3.

The origin of the discrepancy between the NW radii determined from
the scanning electron micrographs and from the modeling of the GISAXS
intensity is in the bundling of NWs. The bundling is almost absent
for sample 1, and the cross-sections of the NWs obtained from the
micrographs characterize the NWs along their full length. As the NWs
grow in length, they bundle together, which causes an apparent radial
growth. Simultaneously, the NW density decreases, so that the fraction
of the surface covered by the NWs remains constant \cite{kaganer16bundling}.
The GISAXS provides a statistics of the NW radii averaged over their
lengths, while the top-view micrographs reveal their distribution
at the top. That results in a progressive difference between the distributions
obtained by the two methods.

The widths of the circularity distributions in the right column of
Fig.~\ref{fig:histograms} slightly reduce with the growth of the
NWs. The NW images in the scanning electron micrographs become more
circular since, during NW growth, the bundled nanowires attain a common
shape that tends to a regular hexagon. Also, the low-circularity wing
of the circularity histogram reduces, because the effective radii
of the bundled NWs increase, and the distinction between the top facets
and the top parts of the side facets becomes more pronounced for the
ImageJ analysis. The circularity distributions obtained from the GISAXS
intensity curves are sharper than the ones obtained from scanning electron
micrographs, because the former takes into account both single NWs
in their bottom part and bundled NWs in their top part, while the
latter counts only the NW tops. We also remind that the circularity
of a distorted hexagon is always smaller than the circularity $C\approx0.907$
of a regular hexagon. Larger circularities obtained from the analysis
of the scanning electron micrographs in Figs.~\ref{fig:histograms}(d)
and ~\ref{fig:histograms}(f) are due to the finite resolution of
the micrographs as well as to the algorithm used by ImageJ that tends
to round faceted objects.

We have seen in Sec.~\ref{subsec:Scattering-prism} and particularly
in Fig.~\ref{fig:stars}(f) that, when the scattering vector is oriented
normal to the side facets of the NWs ($\psi=0\,^{\circ}$) and the
facets are atomically flat, the facet truncation rod scattering would
result in a linear increase of the intensity on the $I(q_{x})q_{x}^{4}$
vs.~$q_{x}$ plot at large $q_{x}$. The decrease of the experimental
curves indicates a roughness of the side facets. We obtain in the
Monte Carlo modeling an rms roughness of $\sigma=0.9$, 0.95, and
0.85~nm for samples 1, 2, and 3, respectively. According to the height
of the atomic steps on a GaN($1\bar{1}00$) facet $d_{0}=a_{0}\sqrt{3}/2=0.276$~nm
(here $a_{0}=0.319$~nm is the GaN lattice spacing), the rms roughness
is less than 3.5 steps.

\section{Discussion and summary}

GaN NWs nucleate spontaneously on Si(111) and grow with a substantial
disorder with respect to their orientations. Their growth is, nevertheless,
epitaxial: the NWs inherit the out-of-of plane and in-plane orientations
of the substrate. Since these NWs are typically long (from hundreds
of nanometers to a few microns) and thin (tens of nanometers), the
range of orientations of their long axes of 3--5\,$^{\circ}$ is
sufficient to provide the same average in the small-angle scattering
intensity as if they would have all orientations. However, an angular
range of orientations of the side facets of 3\,$^{\circ}$ gives
rise to features in the GISAXS intensity distribution that are reminiscent
of the crystal truncation rod scattering from flat surfaces of single
crystals.

We have found that the GISAXS intensity depends on the orientation
of the side facets with respect to the incident X-ray beam direction.
In our experiment, the incident beam is kept normal to the average
direction of the long axes of the NWs. The orientation of the incident
beam with respect to the side facets is varied by rotating the sample
about the substrate surface normal. The scattering intensity is maximum
when the incident beam is along the facets, or in other words, when
the scattering vector is in the direction of the facet normal.

The X-ray scattering intensity from a planar surface is proportional
to $q^{-2}$. Porod's law $q^{-4}$ is a result of a full average
over all orientations of the plane \cite{sinha88}, i.e., the integration
over the two angles defining the plane orientation. For GaN NWs on
Si(111), the range of orientations of the long axes is large enough
to provide an integration over the tilt angle and give rise to a $q^{-3}$
dependence when the scattering vector is along the facet normal. In
the $Iq^{4}$ vs.~$q$ plots shown in Fig.~\ref{fig:Iq4}, this
dependence is seen as a linear increase at $\psi=0\,^{\circ}$ or
$60\,^{\circ}$. The intensity decreases as the sample is rotated
about the normal to the substrate surface. The minimum intensity value
is reached at $\psi=30\,^{\circ}$, i.e., in the direction between
facets.

%-----------------------------------------------------------------

The surface roughness gives rise to a decrease in the intensity at
$q\sigma\gtrsim1$, where $\sigma$ is the rms roughness. The Monte
Carlo modeling of the experimental curves in Fig.~\ref{fig:Iq4}
gives $\sigma$ from 0.85 to 0.95~nm, which is just 3.5 times the
height of the atomic steps. Nevertheless, this small roughness strongly
modifies the intensity curve for high values of $q$.

The GISAXS curves vary fairly little from one sample to another, despite
the large difference between the cross-sectional sizes of the NWs
observed in the scanning electron micrographs shown in Fig\@.~\ref{fig:SEMimages}.
This apparent discrepancy is explained by the NW bundling, which is
an essential effect in their growth \cite{kaganer16bundling}. While
GISAXS reflects the distribution of the cross-sectional sizes of the
NWs over their whole volume, the top-view micrographs shown in the
right column of Fig.~\ref{fig:SEMimages} reveals the cross-sectional
sizes of the NWs at their very top part. As a result, the distributions
of the NW radii and circularities obtained from the scanning electron
micrographs and the GISAXS intensity curves only coincide for sample
1, which is free of bundling. As the NWs grow in height and their
bundling increases (samples 2 and 3), the discrepancies between the
results obtained by these two different methods increases.

Finally, we conclude that GISAXS, together with the Monte Carlo modeling
of the intensity curves, is well suited for the determination of the
distributions of the cross-sectional sizes of the NWs. The methods
developed in the present paper are not specific to GaN NWs on Si(111)
and can be applied to other NW distributions and material systems.
Particularly, they will be applied in a separate work to the assessment
of the radius distributions of GaN NW ensembles grown on TiN, which
exhibit a much lower density that hinders the analysis of the NW cross-sectional
shapes by scanning electron microscopy.

\section{Acknowledgments}

We thank Vladimir Volkov (Institute of Crystallography, Moscow) for
useful discussions, and Oliver Brandt (Paul-Drude-Insitut, Berlin)
for useful discussions and a critical reading of the manuscript, Lewis
Sharpnack (ESRF, Grenoble) for his assistance during preliminary measurements
at the ESRF beamline ID02, Shyjumon Ibrahimkutty (Rigaku, Neu-Isenburg)
for the measurement of the NW twist angles, Carsten Stemmler (Paul-Drude-Institut,
Berlin) for his help with the preparation of the samples, and Anne-Kathrin
Bluhm (Paul-Drude-Institut, Berlin) for providing the scanning electron
micrographs. S.F.-G. acknowledges the partial financial support received
through the Spanish program Ramón y Cajal (co-financed by the European
Social Fund) under grant RYC-2016-19509 from the former Ministerio
de Ciencia, Innovación y Universidades.

%\referencelist[surface]

%\bibliographystyle{plain}
%\bibliography{surface}

\end{document}